\documentclass{iaus}
\usepackage{graphicx}

\newcommand{\HI}{{\sc HI}}
\newcommand{\msun}{$M_\odot$}

\title[The Evolution of LCBGs:  Disks or Spheroids?] %% give here short title %%
{The Evolution of Luminous Compact Blue Galaxies:  Disks or Spheroids?}

\author[D.J. Pisano et al.]   %% give here short author list %%
{D.J. Pisano$^1$, K. Rabidoux$^1$, C.A. Garland$^2$, R. Guzm\'an$^3$, F.J. Castander$^4$, \and J. P\'erez-Gallego$^3$}

\affiliation{$^1$West Virginia University Dept. of Physics, P.O. Box 6315, Morgantown, WV 26506, USA
 \\ email: {\tt djpisano@mail.wvu.edu, krabidou@mix.wvu.edu} \\[\affilskip]
$^2$Natural Sciences Department, Castleton State College, Castleton, VT 05735, USA \\email: {\tt catherine.garland@castleton.edu}\\[\affilskip]
$^3$Dept. of Astronomy, University of Florida, 211 Bryant Space Science Center, P.O. Box 112055, Gainesville, FL 32611, USA
\\email: {\tt guzman@astro.ufl.edu, jgallego@astro.ufl.edu}\\[\affilskip]
$^4$Institut de Ci\'encies de l'Espai (ICE/CSIC), Campus UAB, 08193 Bellaterra, Barcelona, Spain \\email : {\tt fjc@ieec.fcr.es}}

\pubyear{2011}
\volume{277}  %% insert here IAU Symposium No.
\pagerange{1--4}
\jname{Tracing the Ancestry of Galaxies on the Land of our Ancestors}
\editors{C. Carignan, K.C. Freeman, \& F. Combes, eds.}
\begin{document}

\maketitle

\begin{abstract}
Luminous compact blue galaxies (LCBGs) are a diverse class of galaxies characterized by high luminosity, blue color, and high surface brightness that 
sit at the critical juncture of galaxies evolving from the blue to the red sequence.  As part of our multi-wavelength survey of local LCBGs, we have been 
studying the \HI\ content of these galaxies using both single-dish telescopes and interferometers.  Our goals are to determine if single-dish \HI\ 
observations represent a true measure of the dynamical mass of LCBGs and to look for signatures of recent interactions that may be triggering star 
formation in LCBGs. Our data show that while some LCBGs are undergoing interactions, many appear isolated. While all LCBGs contain \HI\ and 
show signatures of rotation, the population does not lie on the Tully-Fisher relation nor can it evolve onto it. Furthermore, the \HI\ maps of many LCBGs 
show signatures of dynamically hot components, suggesting that we are seeing the formation of a thick disk or spheroid in at least some LCBGs. There 
is good agreement between the \HI\ and H$\alpha$ kinematics for LCBGs, and both are similar in appearance to the H$\alpha$ kinematics of high 
redshift star-forming galaxies. Our combined data suggest that star formation in LCBGs is primarily quenched by virial heating, consistent with model 
predictions. 
\end{abstract}

\firstsection % if your document starts with a section,
\section{Introduction}

When the universe was 4.6 Gyr old, the galaxy population was dominated by blue, star-forming galaxies.  Up to 40\% of these galaxies
were luminous compact blue galaxies (LCBGs) which contribute significantly to the global star formation rate density at that time \cite{guzman97}.  
Today, the population of galaxies is roughly evenly divided between a red and a blue population and the star formation rate density has dropped by
an order of magnitude.  Similarly, LCBGs are an order of magnitude less common \cite{werk04} and contribute negligibly to the global star 
formation rate \cite{guzman97}.  LCBGs are a diverse class of galaxies characterized by their high luminosities (M$_B \le$-18.5 mag), compact sizes (SBe(B)$\le$21 mag 
arcsec$^{-2}$, equivalent to r$_{eff}\le$4 kpc), and blue colors ($B-V \le$0.6 mag); they
have the highest star formation rate per unit mass for high mass galaxies \cite{gildepaz00}.  Typical stellar masses of LCBGs are $\sim$5$\times$10$^{10}$\msun, 
\cite{guzman03} placing them near the maximal stellar mass of the blue sequence \cite{kauffmann03}.  Above this mass limit, all galaxies are red so some 
process must quench the star formation in galaxies as they grow.  There have been numerous theories as to what quenching mechanisms operate in galaxies 
on the blue sequence.  These include the shock heating of gas to the virial temperature \cite[(Cattaneo et al. 2006)]{cattaneo06}, or heating by starbursts 
by supernovae- or AGN-driven winds or some combination of multiple processes \cite[(Hopkins et al. 2006, and references therein)]{hopkins06}.  
Since LCBGs reside at the high mass end of the blue sequence, they are poised to have their star formation quenched in the
near future and, therefore, represent an ideal population to study viable quenching mechanisms that could also be responsible for the emergence
of a red sequence in the past 8 Gyr.  

\begin{figure}[b]
\begin{center}
\includegraphics[width=0.4\textwidth]{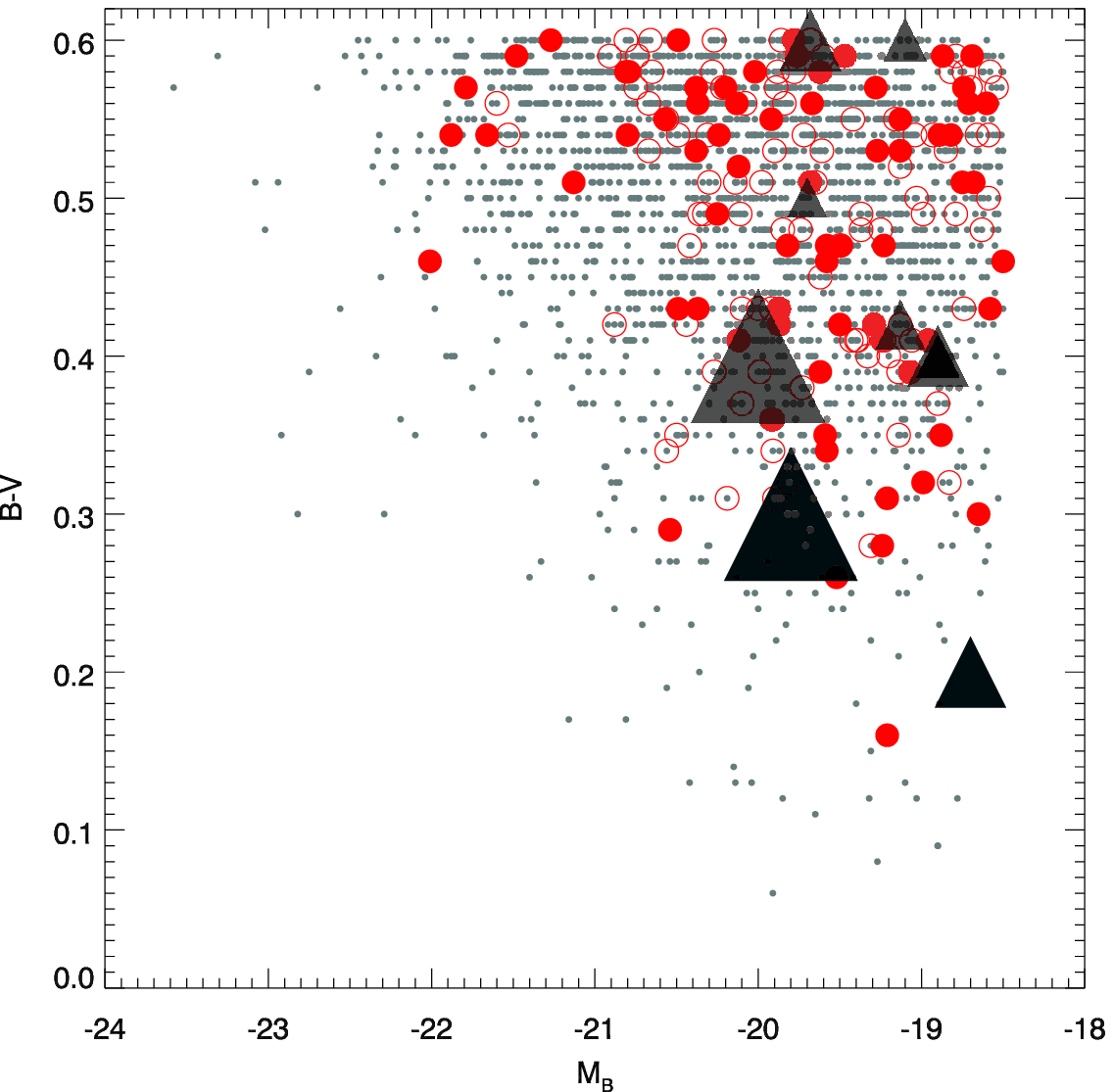}
\includegraphics[width=0.4\textwidth]{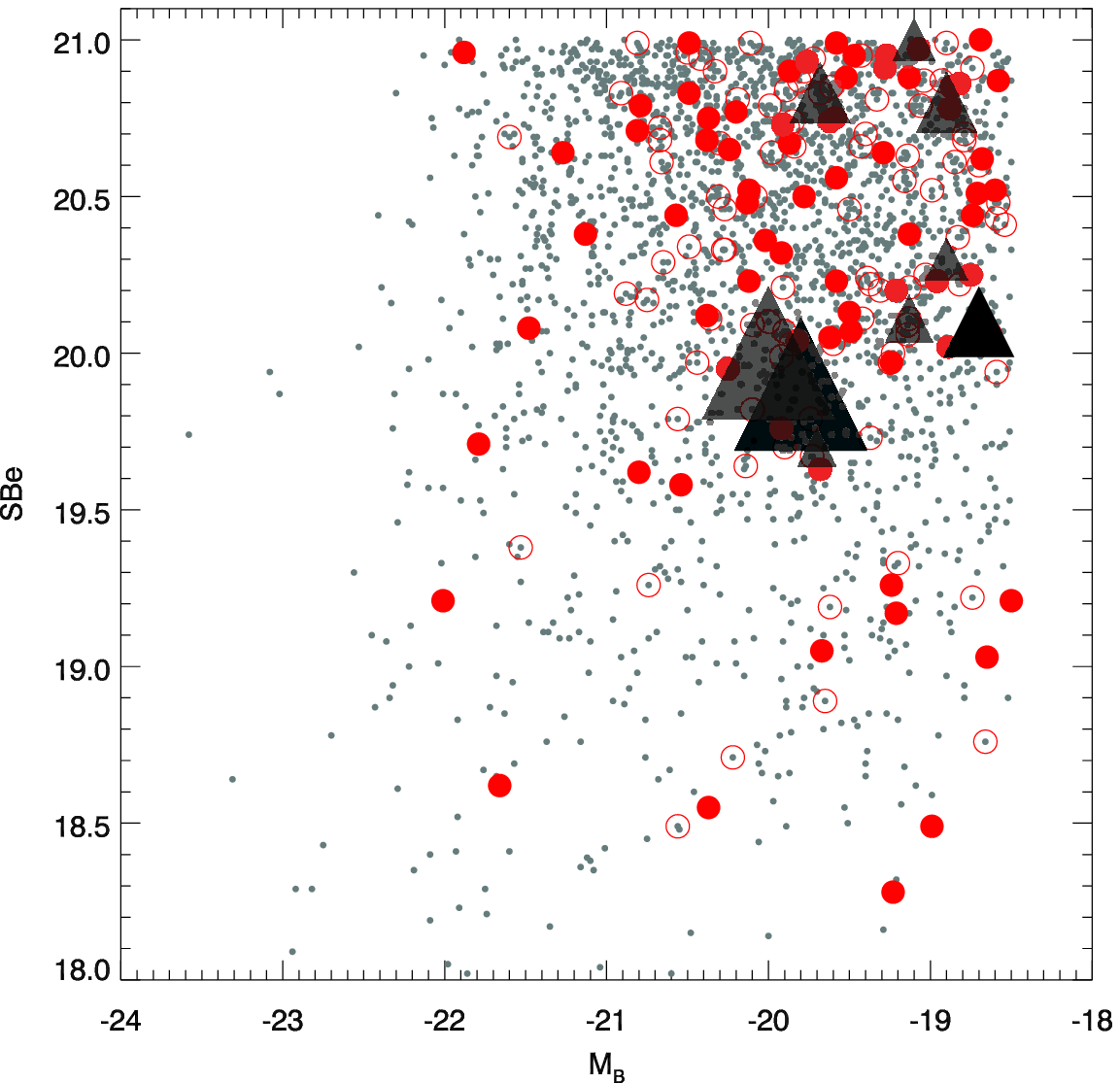}
 \caption{Left:  $B-V$ vs. M$_B$ for all LCBGs selected from the SDSS DR4 (grey dots) and all LCBGs with single-dish \HI\ data (circles).  The filled
 circles are those LCBGs with companions within the beam, the open circles are isolated.  The triangles represent those LCBGs with GMRT or VLA
 data with the size of the triangle inversely proportional to V$_{rot}$/$\sigma$.  Right:  Same as the left but for SBe(B) vs. M$_B$.}
   \label{fig1}
\end{center}
\end{figure}

We are conducting a multi-wavelength survey, spanning the ultraviolet through the radio, of the rare, local LCBGs to constrain the viable mechanisms for
quenching star formation in blue galaxies and the future evolutionary paths of LCBGs.  Therefore, we have selected our LCBGs from the Sloan Digital Sky 
Survey (SDSS) within D$\le$200 Mpc to have the same properties, listed above, as LCBGs at high redshift.  This yields a 
total of 2359 LCBGs out of over 800,000 galaxies in the SDSS DR4.  Of these, we have collected single-dish \HI\ observations of 163 LCBGs.  The distribution of properties for all 
LCBGs are shown in Figure~\ref{fig1}.  

\section{Results}

Single-dish \HI\ observations permit a direct measure of the amount of fuel available for star formation, $M_{HI}$, and, by using the linewidth
and an estimate of a galaxy's extent, the dynamical mass, $M_{dyn}$.  When combined with a measure of the star formation rate, such as
the non-thermal radio continuum emission or emission from dust in the far-infrared, this provides an estimate of the duration of the current starburst:
$\tau = M_{HI}/SFR$.  

We have used original observations and archival data from Arecibo, the Green Bank Telescope, Parkes, Nan\c{c}ay, and the old Green Bank 
140-foot and 300-foot telescopes \cite[(Springob et al. 2005, Giovanelli et al. 2005)]{springob05,giovanelli05} in concert with radio continuum fluxes from NVSS 
and far-infrared fluxes from IRAS to measure $M_{HI}$, $M_{dyn}$, and $\tau$ for 163 LCBGs.  Of the 63 galaxies we observed ourselves, we detected 94\% of 
them at the 5$\sigma$ level of 2.5$\times$10$^8$\msun.  LCBGs have a wide range of properties with an average $M_{HI}\sim$10$^{9.7}$\msun\ and an average
$M_{dyn}\sim$10$^{10.6}$\msun; 80\% $\tau\le$3 Gyr \cite[(Garland et al. 2004, Pisano et al. 2011, in preparation)]{garland04}.  All LCBGs have
$M_{dyn}\le$10$^{12}$\msun, below the maximal halo mass predicted by \cite{cattaneo06}, suggesting that virial heating is a viable quenching mechanism.  Finally, 
there is no significant difference between the \HI\ properties of LCBGs with and without nearby companions; about 50\% of LCBGs have close optically-bright 
companions.  These properties suggest that LCBGs may evolve into low-mass spiral galaxies or high-mass dwarf ellipticals or, possibly, bulges.  Given the wide range 
of properties of LCBGs, the specific evolutionary path of a galaxy depends critically on its exact properties, such that each LCBG will follow a different evolutionary path.  

There is, however, a caveat to this relatively simple evolutionary picture.  LCBGs lie well above and below the Tully-Fisher relation, so many cannot even evolve onto it.  This may be due to nearby companions confusing measurements of the integrated \HI\ linewidth or from significant deviations from pure circular rotation in the LCBGs
due to recent interactions.  In order to quantify these effects, we have mapped a total of 18 LCBGs with the VLA or the GMRT.  Initial analysis for three interacting 
LCBGs was presented in \cite{garland07}, and we present the results for eight additional LCBGs here.  

\begin{figure}[th]
\begin{center}
\includegraphics[width=0.8\textwidth]{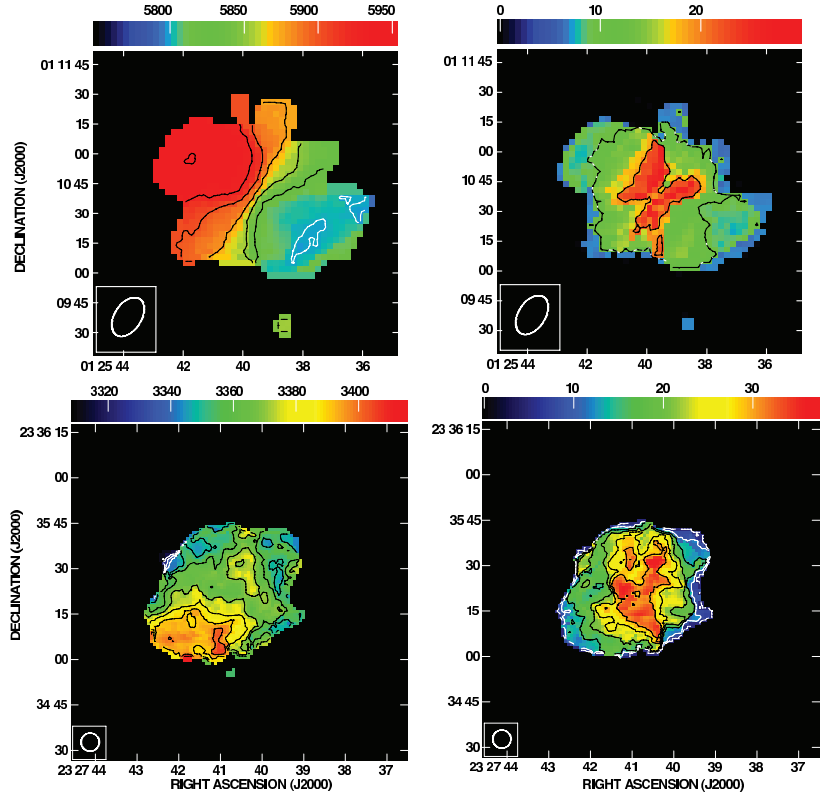}
\caption{A montage of velocity fields (left column) and maps of velocity dispersions (right column) for SDSS~0125+0110 (top row) and Mrk~325 
(bottom row).  Velocity field contours are every 25 km s$^{-1}$ for SDSS~0125+0110 and every 10 km s$^{-1}$ for Mrk~325.  Contours on the
velocity dispersion maps are every 10 km s$^{-1}$ and every 5 km s$^{-1}$ for the two galaxies, respectively.  The beam size is shown in the lower left corner of
each panel.}
\label{fig2}
\end{center}
\end{figure}

As can be seen from Figure~\ref{fig2}, while some LCBGs have regular rotation and relatively small velocity dispersions others have a far more complex velocity
structure, even if they lack a close companion.  Our high resolution GMRT observations reveal that the single-dish linewidths are about 15\% larger than what would be 
inferred from the interferometer data.  This overestimate of the rotation velocity is not large enough, however, to explain the dispersion seen around the Tully-Fisher
relation.  

Our observations also show that LCBGs tend to have high velocity dispersions, $\sim$20-40 km s$^{-1}$, across much of the galactic disk.  For those LCBGs with 
large rotation velocities, this leads to $V_{rot}/\sigma\sim$5-7 (seen in the top row of Figure~\ref{fig2}), however many LCBGs have much smaller  values of 
$V_{rot}/\sigma$ (seen in the bottom row of Figure~\ref{fig2}) of $\lesssim$2.  Similar features are seen in optical velocity fields of star-forming high-redshift galaxies
\cite[(Shapiro et al. 2008, Gon{\c c}alves et al. 2010, and these proceedings)]{shapiro08,goncalves10}.  Such low $V_{rot}/\sigma$ values indicate the presence of a 
dynamically hot component in a number of LCBGs, perhaps indicating the formation of a thick disk or a bulge after a recent minor merger.  Our \HI\ velocity fields agree 
with the H$\alpha$ velocity fields observed by \cite{perez-gallego11}, suggesting that the dynamics of the ionized gas are not severely affected
by galactic outflows.  

\section{Conclusions}

The signature of ongoing spheroid formation in some LCBGs is consistent with the idea that star formation in these galaxies is being quenched via virial heating, but
this is not a unique explanation.  Figure~\ref{fig1} shows that those LCBGs with the smallest values of $V_{rot}/\sigma$ are the most compact, bluest, and highest
luminosity systems.  This could also indicate that quenching from heating due to the intense central starburst or its associated supernovae is a possibility.  This is 
supported by the results of optical spectroscopy by \cite{perez-gallego11} who found that while only 5\% of LCBGs have an AGN, 27\% have signatures of supernovae-
driven winds.  The remaining LCBGs could then be quenched via virial heating.  In the future, we will be expanding our \HI\ mapping to study additional LCBGs with a 
wider range of properties and we will use multi-wavelength data to search for signatures of active quenching in LCBGs.

%\begin{discussion}

%\end{discussion}

\end{document}